# Identification of Utility-Scale Renewable Energy Penetration Threshold in a Dynamic Setting

Hashem Albhrani , *Student Member, IEEE*, Reetam Sen Biswas, *Student Member, IEEE*, and Anamitra Pal, *Senior Member, IEEE*

School of Electrical, Computer, and Energy Engineering
Arizona State University, Tempe, Arizona-85287, USA
Email: halbhran@asu.edu, rsenbisw@asu.edu, anamitra.pal@asu.edu

*Abstract*— Integration of renewable energy resources with the electric grid is necessary for a sustainable energy future. However, increased penetration of inverter based resources (IBRs) reduce grid inertia, which might then compromise power system reliability. Therefore, power utilities are often interested in identifying the maximum IBR penetration limit for their system. The proposed research presents a methodology to identify the IBR penetration threshold beyond which voltage, frequency, and tie-line limits will be exceeded. The sensitivity of the IBR penetration threshold to momentary cessation due to low voltages, transmission versus distribution connected solar generation, and stalling of induction motors are also analyzed. Dynamic simulation studies conducted on a 24,000-bus model of the Western Interconnection (WI) demonstrate the practicality of the proposed approach.

## I. INTRODUCTION

If we continue to use conventional fossil-fuel based energy resources (e.g., coal, oil, and natural gas) in an unabated fashion, it is only a matter of time before they get exhausted [1]. Additionally, the increased use of fossil fuels pollutes the environment and accelerates global warming. As such, to create a path for a greener and more sustainable energy future, there is a strong push for increasing the penetration of renewable generation in the modern power system [2]. Figure 1 shows the steady increase in penetration of renewable generation resources in the US electric grid [3]. However, from the perspective of reliability, a power system with high penetration of renewable energy resources is more difficult to manage [4]-[5].

There are primarily two challenges that power utilities face with regards to high penetration of renewable energy. First, the system inertia reduces when synchronous generators are replaced with inverter based resources (IBRs), which in turn, affects the frequency stability [6] and transient stability of the system [7]. Due to the limited fault-ride through capabilities of IBRs, voltage stability of the system is also compromised [8]. Recent research has tried to address this challenge by developing synthetic inertia (SI)-based power electronic controllers; for instance, in [9], SI controllers provided fast frequency response during system transients. The second challenge is the uncertainty associated with the renewable energy resources, because of the variability of their input sources, namely, solar irradiance and wind speed. To account for the uncertainty and intermittency of IBRs, the state-of-the-art research has often focused on creating *buffers* in the form of battery energy storage systems (BESSs) for ensuring continued reliability.

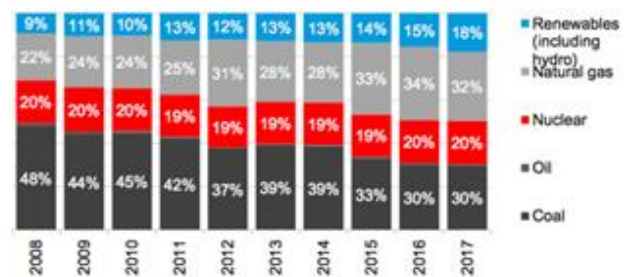
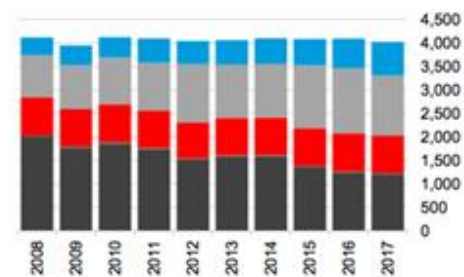

Figure 1: Energy generated from different sources in the US through the years 2008-2017 in percentages and terawatt-hour (TWh) [3].

Considering the above-mentioned challenges associated with IBRs, power utilities are often interested in finding the maximum threshold of IBR penetration in their systems beyond which the reliable operation of their bulk power system will be compromised, and they will need interventions in the form of SI-based controllers and/or BESSs. Research has already been done in simulating high renewable generation scenarios using various simulation programs [10]-[12]. In [10], power system transient analysis was performed in a case that had the synchronous generators replaced with photovoltaic (PV) units using MATLAB/Simulink. In [11], low frequency oscillations were identified with transient security analysis (TSA) tools and analyzed using the Prony algorithm to identify the locations where IBRs can be installed. Reference [12] performed dynamic contingency analysis in presence of renewable generation using PSLF and addressed the risks of integrating intermittent renewable resources into the power grid.

However, there is still scope for more research to be done in this area. For example, the effects of increased penetration of IBRs on the tie-line power flows, especially during contingencies, has not received significant attention.

Similarly, the effects of different types of systemic and/or operational directives on the IBR penetration threshold must be explored in more details. Keeping these two aspects in mind, the research done in this paper makes two contributions to the state-of-the-art:

(1) It finds the maximum IBR penetration threshold of a power utility in the Western Interconnection (WI), with regards to both WECC TPL criteria [13], and tie-line power transfer limits [14], and

(2) It investigates how the IBR penetration threshold changes with regards to three different sensitivities, namely, momentary cessation due to low voltages, transmission versus distribution connected solar generation, and stalling of induction motors.

In summary, this paper presents a methodology to determine the ability of IBR-rich power grids to handle different system conditions.

The rest of the paper is structured as follows. Section II explains the approach developed to find out suitable IBR penetration threshold beyond which voltage, frequency, and tie-line limits will be exceeded. Section III discusses the effect of different sensitivities on the IBR penetration thresholds computed in Section II. Finally, Section IV presents the concluding statements.

## II. Proposed Approach for Identifying IBR Penetration Threshold

### A. Database

A power flow and a dynamic file of the WI was provided by the utility to find the maximum IBR penetration threshold that their system can reliably handle. The power flow file had approximately 24,000 buses, 4,300 generators, 12,000 loads and 18,500 transmission lines. The dynamic file had information about the models for non-renewable as well as renewable entities (generators, loads) pre-existing in the system. A dynamic simulation lasting 20 seconds was run in PSLF initially using the given power flow and dynamic files to ensure correct initialization of the dynamic models. 56 contingencies (48 transmission line contingencies and 8 generator contingencies) were also provided by the utility to analyze system performance. A user defined EPCL script was written to process these contingencies in PSLF.

### B. Criteria for analysis

The WECC TPL criteria for voltage and frequency [13], [14], and the tie-line power transfer limits provided by the utility were used as metrics to identify exceedances during the dynamic simulation. The details of the WECC TPL criteria are summarized below.

1. *59.6 Hz frequency deviation criteria*: The frequency at any bus cannot remain below 59.6 Hz for more than 6 cycles, as shown in Figure 2.

2. *WECC voltage recovery criteria*: After the fault has been cleared at a bus, the voltage must recover to 80% of its initial voltage within 20 seconds as shown in Figure 3.

3. *WECC 70% voltage dip criteria*: The time duration of the voltage dip below 70% of the initial voltage must not be more than 30 cycles.

4. *WECC 80% voltage dip criteria*: The time duration of the voltage dip below 80% of the initial voltage must not be more than 2 seconds. Figure 4 depicts both the 70% voltage dip criteria and the 80% voltage dip criteria.

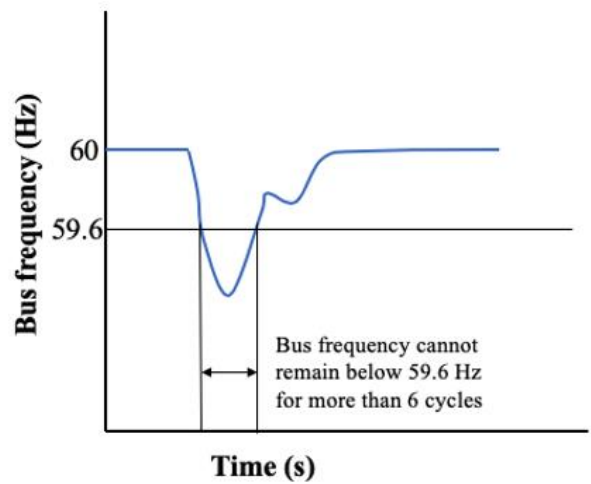

Figure 2: An example of frequency going below 59.6 Hz and how the duration of the recovery is calculated [14]

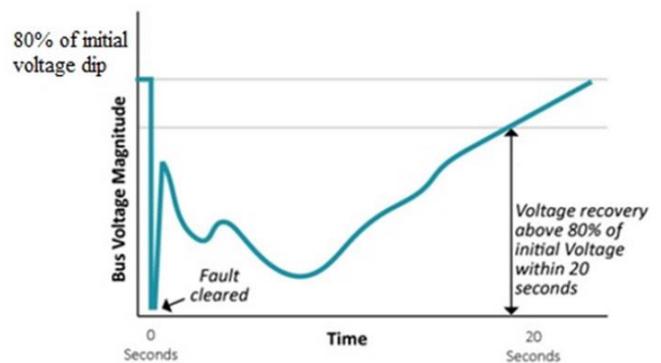

Figure 3: An example of voltage going below the original value and how the duration of recovery is calculated [14]

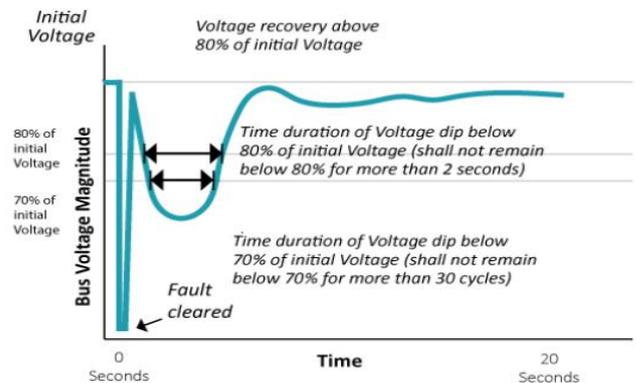

Figure 4: An example of voltage going below the original value and how the duration of recovery is calculated [14]

In absence of SI based controllers or BESSs, a very high IBR penetration in the area-under-study will reduce the frequency and voltage response capabilities of that region significantly [9], [15]. Under such circumstances, the area will have to rely on its neighbors to provide necessary power to ride through transient events such as generator contingencies. However, neighboring utilities have agreements on how much power can be exported or imported through the tie-lines [16]. As such, power exchange beyond the permissible limit will be deemed a *violation* of the agreement and must be avoided. The "imetr" model in PSLF [17] was used to monitor the tie-line power flows during the time domain simulations. This model was added to all the tie-lines and transformers connecting the area-under-study with its neighboring areas.

Next, the output of the time domain simulations was analyzed to check if it satisfied the above-mentioned criteria. An EPCL script was created to check if the WECC TPL criteria were exceeded. A separate Python script was created to check if the tie-line power flows were within the respective limits throughout the length of the simulation.

*C. Dynamic contingency analysis in original case*

Figure 5 presents the current energy resource mix of the area-under-study. It is observed from the figure that the IBR penetration in the original system is 11%, while natural gas accounts for 41% of the power generation. Even in the original system some contingency cases violated the tie-limit flows by about 60 MVA as shown in Figure 6. It was determined (with approval from the utility) that for IBR-rich system, the goal would be to ensure that the tie-line power flows do not exceed that of the original system (henceforth, termed base-case). It must be noted that the goal here is not to improve the base-case in terms of the number/type of violations, but to ensure that the IBR-rich system does not create additional violations than that present in the base-case.

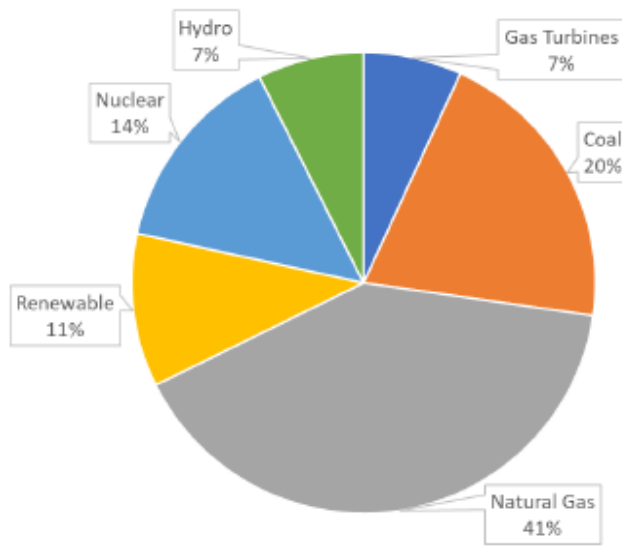

Figure 5: Distribution of power generation in the original system

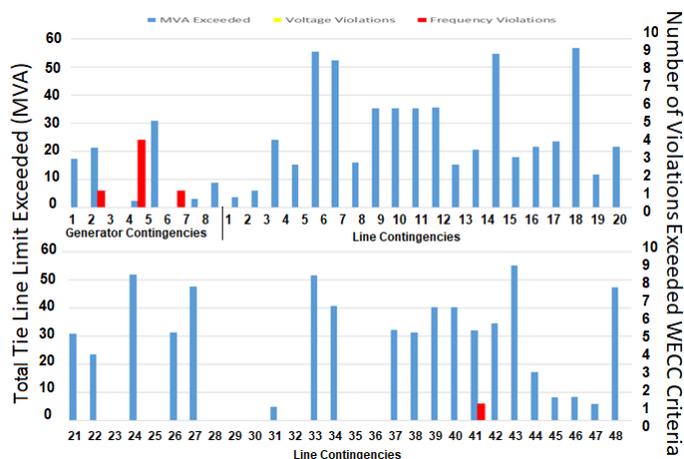

Figure 6: Results of dynamic contingency analysis for the original system

*D. Replacement of synhcronous generation with IBR*

The current trend of going towards "clean energy" not only implies adding IBRs to the power grid, but also to shut down the most polluting sources of energy, such as those based on fossil fuels [18]. Therefore, it was necessary to first identify the synchronous generators in the system that could be replaced before adding new IBRs to the system. In consultation with the utility, coal-fired and gas-turbine units were identified for potential replacement; for the area-under-study, coal-fired and gas-turbine units had a total capacity of approximately 7,400 MW. When all the coal-fired and gas-turbine units were shut down and replaced by IBRs (added at locations provided by the utility), the IBR penetration increased from 11% to 41%.

*E. Dynamic contingency analysis for higher IBR case*

Dynamic contingency analysis was performed on the 41% IBR penetration case created in the previous sub-section. Figure 7 presents the results for the 41% IBR case. The exceedances (especially the tie-line limit violations) had increased significantly in comparison to the base-case (compare Figure 6 with Figure 7). Therefore, iterations were performed between Sections II.D and II.E until similar violations were observed between the base-case and the IBR-rich case. The iterations involved adjustments of the amount of generation of each IBR unit and proportionately restoring some of the synchronous generator units that were removed previously (in Section II.D). Eventually, the IBR penetration threshold for the area-under-study was found to be 28%. Figure 8 shows that the 28% IBR penetration case gave results that were similar to the base-case that contained 11% IBRs (compare Figure 6 with Figure 8). Figure 9 and Figure 10 show the voltage response and frequency response of the 28% IBR penetration case, respectively, for the contingency that had the greatest number of line faults. The plots in Figure 9 and Figure 10 depict that the WECC TPL criteria for voltage and frequency is satisfied by the 28% IBR system.

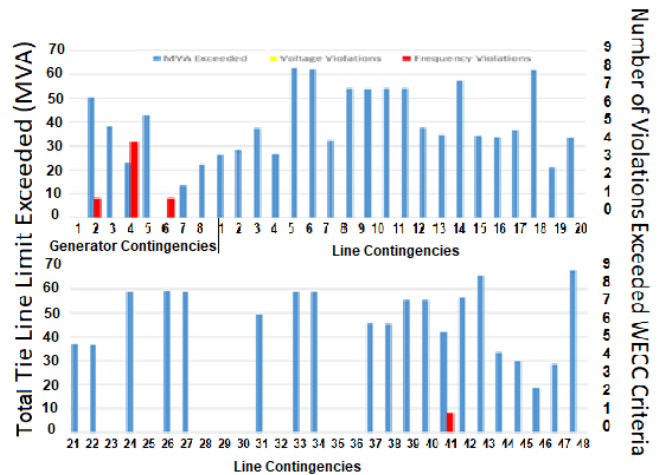

Figure 7: Results of dynamic contingency analysis with 41% IBR generation

### III. EFFECT OF DIFFERENT SENSITIVITIES ON IBR PENETRATION THRESHOLD

Momentary cessation due to low voltages, transmission connected solar generation versus distribution connected solar generation, and stalling of induction motors are three sensitivities that could significantly affect operations of IBR-rich power systems as discussed in [19] and [20]. As such, in this section, we investigate how the IBR penetration threshold computed in Section II for the area-under-study changes with regards to these three sensitivities.

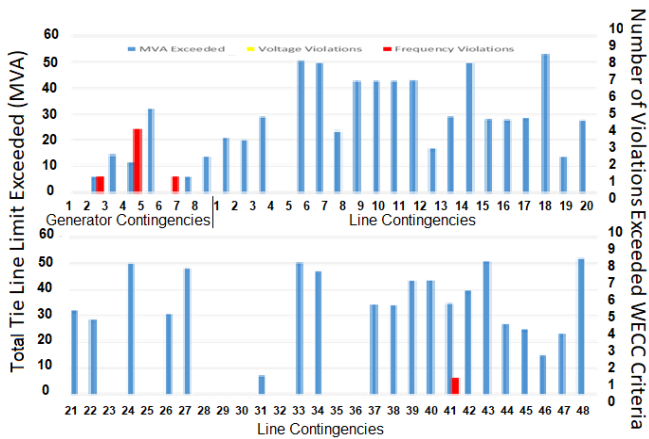

Figure 8: Results of dynamic contingency analysis with 28% IBR generation

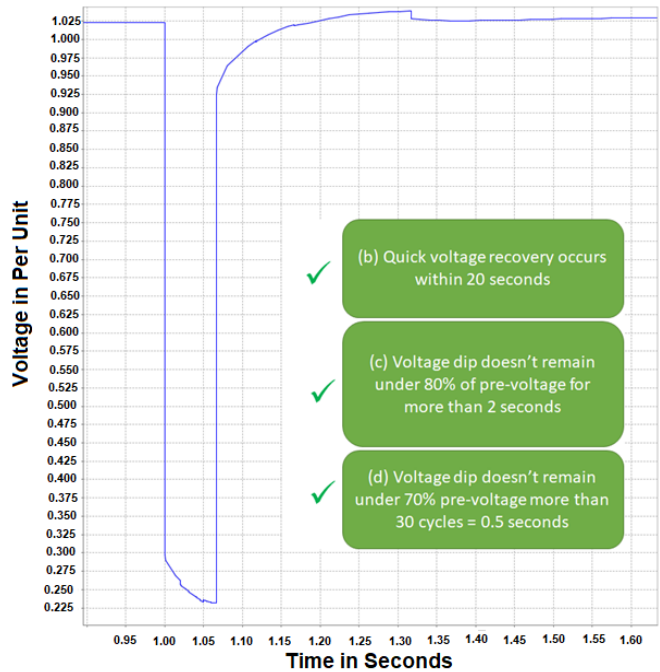

Figure 9: Voltage response for a contingency for 28% IBR case

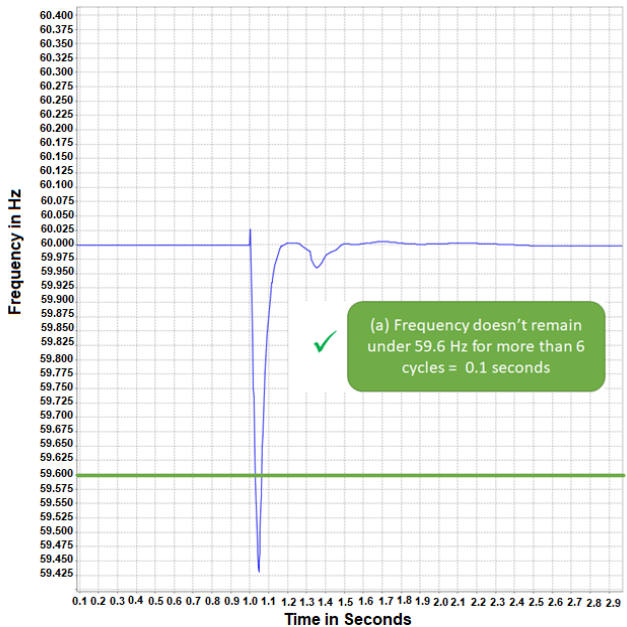

Figure 10: Frequency response for a contingency for 28% IBR case

### A. Momentary Cessation

The need to investigate the reliability of the system with regards to momentary cessation arose after the Blue Cut Fire incident in California [19]. It was a wildfire that had triggered a number of line-faults at different parts of the system, as a result of which the inverters had tripped instantaneously due to low voltages during the fault. Since all the IBRs had ceased to generate power after the fault, this phenomenon was termed "momentary cessation". After investigating this event, NERC recommended the utilities to model momentary cessation of inverters in the planning phase to understand how the system behaves if such a scenario manifests again in the future. As such, momentary cessation is an important criterion that must be considered when planning studies are performed for estimating IBR penetration threshold in the power grid.

In this study, momentary cessation was modeled by changing parameter "zerox" of the low voltage power logic (LVPL) of the "regc_a" model of PSLF [17]. The "zerox" parameter enables momentary cessation, the moment voltage goes below it. As such, momentary cessation will occur as many times as the voltage of the IBR goes below the threshold. In consultation with the utility, the value of the "zerox" parameter was set at 0.4. Figure 11 shows that enabling momentary cessation did not deteriorate the system performance, when compared with the original case[1] (see Figure 8). Therefore, momentary cessation was found to be a non-binding constraint for the area-under-study, and the IBR penetration threshold remained at 28%.

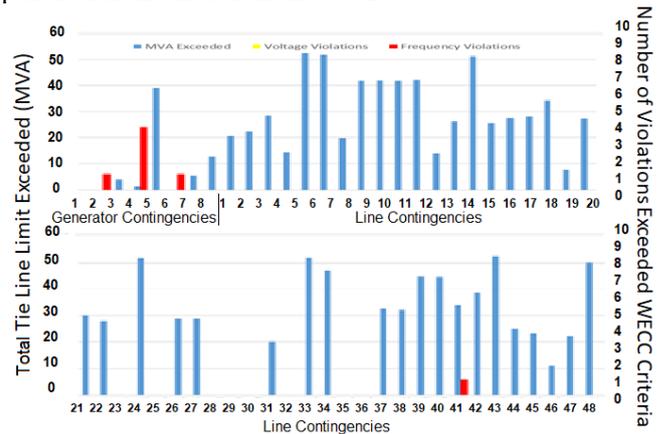

Figure 11: Results of dynamic contingency analysis with 28% IBR generation with momentary cessation enabled

### B. Transmission connected solar generation versus distribution connected solar generation

Another sensitivity study that is valuable for utilities is the difference in the performance of the system when there is transmission connected solar generation versus distribution connected solar generation [21]. This sensitivity analysis can help utilities understand how their system will perform when there is large number of roof-top solar PV generation coming in from the distribution system instead of a solar farm that can be more efficiently monitored and controlled by the utility. To represent the distribution connected solar generation, the "regc_a" model was replaced with the "pvd1" model from PSLF [17]. Figure 12 presents the results with distribution connected solar generation. The graph depicts similar violations with that of the original case[1] in Figure 8. Therefore, the IBR penetration threshold remained at 28%.

---

[1] Note that the original case here refers to the 28% IBR case obtained at the end of Section II.E

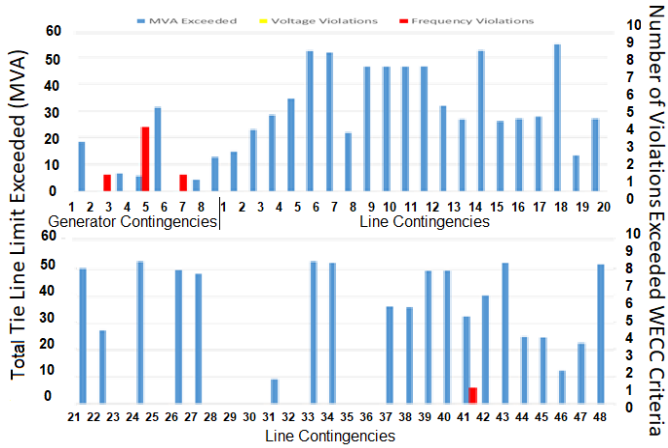

Figure 12: Results of dynamic contingency analysis with 28% IBR generation modeled as being connected at the distribution level

*C. Stalling of induction motors*

The last sensitivity that was considered in this study is the stalling of induction motors, which imitates the stalling of motors of single-phase air conditioners due to low voltages. This sensitivity study involved modifying all the parameter settings of the composite load model "cmpldw" of the region. The first parameter that was changed in the dynamic file was the stall delay time "Tstall"; it was changed from 0.6 p.u. to 0.42 p.u [22]. Next, the threshold voltage for stalling was enabled by changing its value from 9999 p.u. to 0.033 p.u. as mandated by NERC [23]. There were about 800 models in the area-under-study, all of which were modified to correctly simulate the stalling of induction motors.

After completing the needed modifications, the case study was run to check its performance with regards to the WECC TPL criteria and the tie-line limits; the results obtained are shown in Fig. 13. From Figure 13 it can be noticed that the 28% IBR case generated many voltage violations for different contingencies. Therefore, the base-case (see Section II.C) was tested with the stalling of induction motors to check whether it was a pre-existing problem of the system or a new problem that needed to be addressed. It was found that the base-case also exceeded the tie-line limits if the stalling of induction motors was enabled (compare Figure 6 with Figure 14). However, the violations were lesser compared to the 28% case (compare Figure 13 with Figure 14). As the goal of this research is to make a new case with higher IBR penetration that does not create more violations than the base-case, the IBR penetration level was reduced to 15% (from 28%) by following the iterative procedure mentioned in Section II.E. Figure 15 shows that with 15% IBR penetration, the dynamic performance of the system was as good as the base-case (compare Figure 14 with Figure 15).

To summarize, the proposed methodology for estimating IBR penetration threshold for the area-under-study involved three phases as depicted in Figure 16. In the first phase, all coal-fired and gas-turbine units were replaced with IBRs to create a 41% IBR case. This case created significant amount of tie-line limit violations. Hence, the IBR penetration was reduced to 28% in the second phase. Finally, after performing the three different sensitivity studies, stalling of induction motors was found to be a limiting factor that further reduced the IBR penetration threshold to 15% (in the third phase).

## IV. CONCLUSIONS

This paper proposed a systematic methodology to estimate the maximum renewable generation penetration in a power system. It considered WECC TPL criteria and tie-line power transfer limits and performed sensitivity analysis with regards to momentary cessation due to low voltages, transmission versus distribution connected renewable generation and stalling of induction motors to determine the IBR penetration threshold for a power utility in the Western Interconnection. With the above-mentioned constraints considered simultaneously, the IBR penetration threshold was found to be 15% for the area-under-study. It was verified that at this IBR penetration level, the violations were of similar number/type as the original system that was provided by the utility.

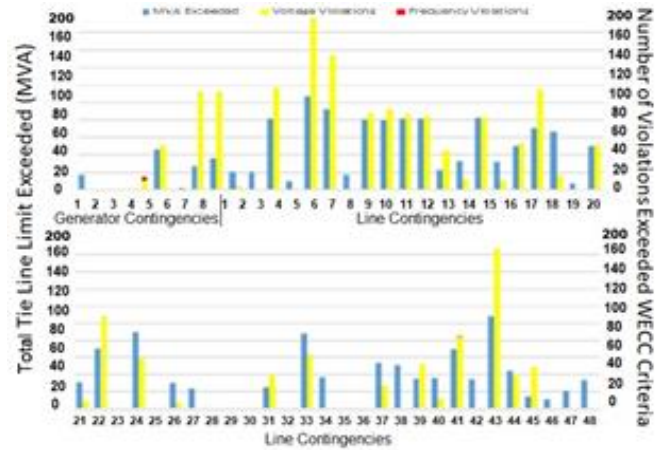

Figure 13: Results of dynamic contingency analysis with 28% IBR case by enabling stalling of induction motors

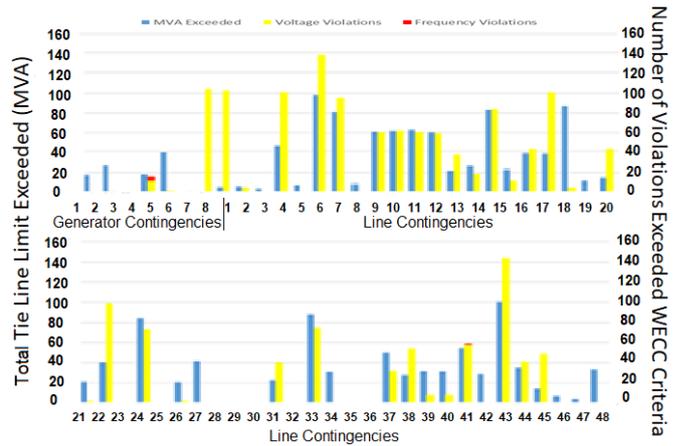

Figure 14: Results of dynamic contingency analysis for original system (11% IBR) by enabling stalling of induction motors

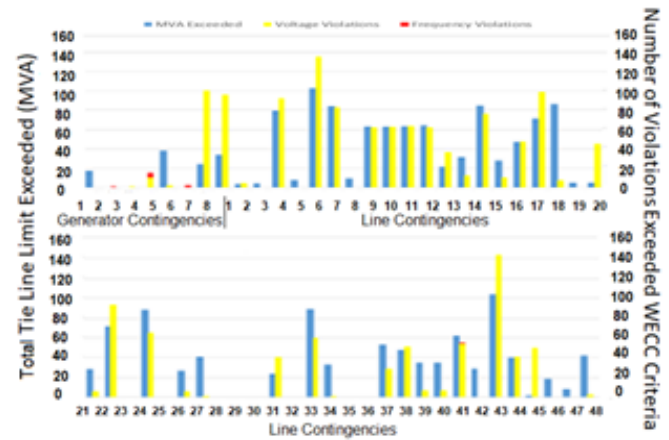

Figure 15: Results of dynamic contingency analysis with 15% IBR case by enabling stalling of induction motors

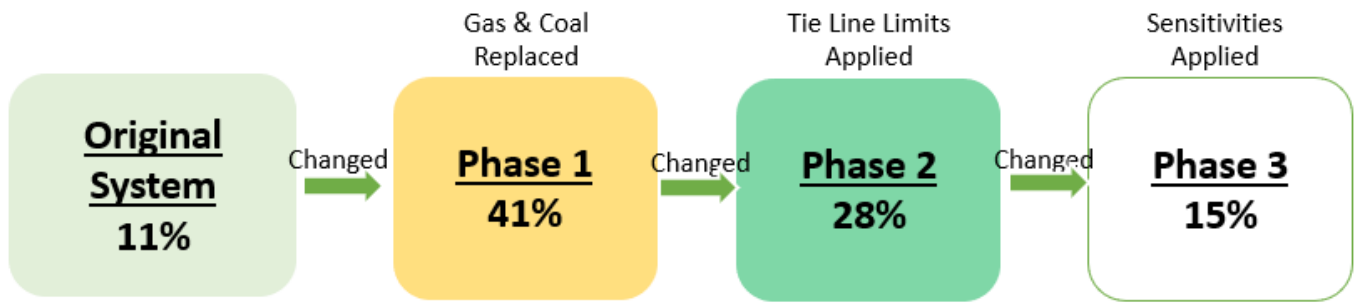

Figure 16: Different phases of the system as the numbers and locations of IBRs and synchronous generator units are changed

The methodology presented in this paper could serve as a guideline to transmission planners of different power utilities to help them come up with a reasonable IBR penetration bound for their respective systems. One reason for the small increase in IBR penetration from the original system (which had 11%) to the new system (which had 15%) for the area-under-study could be that the locations where the IBRs were being added were not optimal. This will be investigated in a future article.